\documentclass[prl,aps,reprint,10pt,nofootinbib,preprintnumbers,superscriptaddress]{revtex4-1}
\pdfoutput=1
\usepackage{geometry}
\usepackage{color}
\usepackage{fullpage}
\usepackage[title]{appendix}
\usepackage{hyperref}
\hypersetup{colorlinks, citecolor=Blue, linkcolor=black, urlcolor=Blue}
\usepackage{amsfonts}
\usepackage{setspace}
\usepackage[usenames, dvipsnames]{xcolor}
\usepackage{tabularx}
\usepackage{empheq}
\usepackage{comment}
\usepackage{enumitem}
\usepackage{slashed}
\usepackage{braket}

\definecolor{light-gray}{gray}{0.9}

\newcommand{\eps}{\epsilon}

\newcommand{\be}{\begin{eqnarray}}
\newcommand{\ee}{\end{eqnarray}}

\newcommand{\beq}{\begin{equation}}
\newcommand{\eeq}{\end{equation}}

\newcommand{\gyy}{g_{a\gamma\gamma}}

\newcommand{\w}{\omega}

\newcommand{\GeV}{\text{GeV}}

\newcommand{\rhoDM}{\rho_{_{\rm DM}}}

\newcommand{\qeff}{q_{\rm eff}}

\newcommand{\vect}[1]{\boldsymbol{#1}}

\newcommand{\pd}{\partial}

\newcommand{\cFF}{c_{_{F^2}}}
\newcommand{\xPT}{\chi \text{PT}}

\definecolor{darkblue}{rgb}{0.0,0.0,0.6}
\definecolor{readableAB}{rgb}{0.7,0,0.7}
\definecolor{colorRTD}{rgb}{.2,.2,.7}

\begin{document}

\bigskip\
\preprint{CERN-TH-2023-136}
\preprint{DESY-23-097}

\title{Quadratic Coupling of the Axion to Photons}

\author{Carl~Beadle}
\author{Sebastian~A.~R.~Ellis}
\affiliation{D\'epartment de Physique Th\'eorique, Universit\'e de Gen\`eve,\\
24 quai Ernest Ansermet, 1211 Gen\`eve 4, Switzerland}
\author{J\'er\'emie~Quevillon}
\affiliation{Laboratoire de Physique Subatomique et de Cosmologie, Universit\'e Grenoble-Alpes, CNRS/IN2P3, Grenoble INP, 38000 Grenoble, France}
\affiliation{CERN, Theoretical Physics Department, 1211 Geneva 23, Switzerland}
\author{Pham~Ngoc~Hoa~Vuong}
\affiliation{Deutsches Elektronen-Synchrotron DESY, Notkestr. 85, 22607 Hamburg, Germany}

\begin{abstract}
We show that the QCD axion couples to the electromagnetic kinetic term at one loop. The result is that if axions make up dark matter, they induce temporal variation of the fine structure constant $\alpha$, which is severely constrained. We recast these constraints on the QCD axion parameter space. We also discuss how to generalise our finding to axion-like particles, and the resulting constraints.
\end{abstract}

\maketitle


\textbf{\emph{Introduction.}} --- The axion is a well-motivated extension of the Standard Model (SM). 
The ``QCD axion'' was originally introduced to explain 
the non-detection of the electric dipole moment (EDM) of the neutron~\cite{Peccei:1977hh, Weinberg:1977ma, Wilczek:1977pj}. Axions have since garnered much interest as a candidate for dark matter (DM)~\cite{Preskill:1982cy,Abbott:1982af,Dine:1982ah}.
Other generic pseudo-scalar particles with an ultraviolet (UV) shift symmetry 
which do not relax the neutron EDM to zero
are also natural DM candidates, and  are typically referred to as ``axion-like particles'' (ALPs).\footnote{We will refer to an ``axion'' when a statement applies to both the QCD axion and an ALP.}
Many experimental searches are directed at discovering an axion, many of which assume it to make up all of the dark matter of the universe. Many of these searches rely on the coupling between the axion and photons~\cite{Sikivie:1983ip},
\begin{align}
    \label{eq:AxionPhoton}
    \mathcal{L} \supset -\frac{\gyy}{4} a F_{\mu\nu}\tilde{F}^{\mu\nu}\ ,
\end{align}
to generate observable signals from axion-photon conversion. This interaction and that of the QCD axion with nuclei are the subject of extensive experimental and theoretical work (see e.g.~\cite{Irastorza:2018dyq,Adams:2022pbo} for recent reviews). 

The axion field, being odd under parity-conjugation (P) and charge-parity-conjugation (CP), does not couple to the kinetic term of the photon and therefore does not lead to a shift in the fine-structure constant $\alpha$ to leading order. Likewise, as we show in the Supplemental Material (\textit{s.m.}) through simple helicity arguments, symmetry prevents the operator of Eq.~\eqref{eq:AxionPhoton} from generating a quadratic axion-photon amplitude.
However, we show for the QCD axion, and generalise to ALPs, that an operator of the form
\begin{align}
    \label{eq:QuadAxionPhoton}
    \mathcal{L}_{a^2F^2} \supset  \cFF\,\frac{\alpha}{16\pi^2} \left(\frac{a}{f_a} \right)^2 F_{\mu\nu} F^{\mu\nu} \ ,
\end{align}
is generated at one loop. 
This operator does not respect the UV shift symmetry of the axion and originates in dynamics that are explicitly symmetry-breaking, with $\cFF$ encoding the origin. 
In the case of the QCD axion, $\cFF$ arises from the same dynamics that generates the potential, which preserves a discrete $\mathbb{Z}_n$ shift symmetry for $a$, such that $\cFF \sim \mathcal{O}(10^{-1})$.
For ALPs, we present two constructions that lead to non-zero $\cFF$, one QCD-like and one invoking an explicit symmetry-breaking operator. In the latter, $\cFF$ directly depends on the explicit symmetry breaking parameter, emphasising the fact that the quadratic operator only exists when the axion shift symmetry is broken.

The operator of Eq.~\eqref{eq:QuadAxionPhoton} leads to time-variation of the fine-structure constant $\alpha$ if the axion has a time-varying field value, as expected for DM axions:
\begin{align}
    \label{eq:AlphaShift}
    \alpha(t) \simeq \alpha\left(1 + \cFF \frac{\alpha}{4\pi^2}\left(\frac{a(t)}{ f_a}\right)^2 \right) \ .
\end{align}
Such a variation in the fine-structure constant is severely constrained by cosmology and experiment~\cite{Olive:2007aj, Stadnik:2015kia, Antypas:2022asj}, and is currently the subject of an intense experimental program (see, e.g.,~\cite{Antypas:2022asj} for a recent review). We demonstrate that these constraints also apply to axions, QCD or otherwise. In particular, we consider constraints from cosmology, violations of the weak equivalence principle, and direct searches for ultralight dark matter. Our results are summarised in Fig.~\ref{fig:QCDplot} for the QCD axion, and in Fig.~\ref{fig:ALPplot} for ALPs.


\vspace{3pt}
\textbf{\emph{Generating the quadratic axion-photon coupling.}} --- In the standard lore, the shift symmetry of the axion implies that a basis can be found such that it is derivatively-coupled to SM fields. In this case, the naive expectation is that the first order at which a quadratic axion-photon coupling is generated will be $\mathcal{O}((\pd_\mu a)^2\,f_a^{-4})$, and will therefore be vanishingly small. However, since axions have a small mass due to a breaking of the shift symmetry, a much larger quadratic axion-photon coupling can be generated. Below we will explore this coupling, first for the QCD axion, and subsequently for an ALP.

\vspace{3pt}
\emph{The QCD Axion.} --- The coupling of the QCD axion to SM fields can be consistently treated in Chiral Perturbation Theory ($\xPT$) associated to the breaking of the approximate $SU(N_f)_L \times SU(N_f)_R$ flavour symmetry of the $N_f$ light SM quarks.\footnote{We will take $N_f=2$ for simplicity, but our results hold for $N_f=3$.} Our guide to understanding the coupling of axions to SM fields is then the neutral pion, which shares the same quantum numbers as the axion.\footnote{In a particular axion coupling parameterisation, there is tree-level mixing between $a$ and $\pi^0$. Since observables should not be parameterisation-dependent, we should already conclude that the axion will have all the same couplings as a $\pi^0$. Due to its transformation properties under the chiral symmetry, the $\eta^{(\prime)}$ is an even better guide, and also possesses a quadratic coupling to photons~\cite{Brodsky:1981rp}.} 

In $\xPT$, the first order at which an operator appears leading to a tree-level coupling of neutral pions to $F^2 \equiv F_{\mu\nu}F^{\mu\nu}$ is $\mathcal{O}(p^6)$. However, the process $\gamma\gamma \to \pi^0 \pi^0$ is experimentally observed to have a cross-section that is only $\sim 10^2$ smaller than that of $\gamma \gamma \to \pi^+ \pi^-$, a tree-level $\mathcal{O}(p^2)$ effect, at $\sqrt{s} \sim 0.4\,\text{GeV}$~\cite{Boyer:1990vu,CrystalBall:1990oiv}. In $\chi \text{PT}$, the large $\gamma\gamma \to \pi^0 \pi^0$ cross-section is explained by the observation that unitarity requires it to be generated at one-loop order involving $\mathcal{O}(p^2)$ operators, and is thus $\mathcal{O}(p^4)$ in the $
\xPT$ power-counting. Importantly, as there is no tree-level $(\pi^0)^2F^2$ operator in the $ \mathcal{O}(p^4)$ $\xPT$ Lagrangian, there can be no counterterm and the amplitude for $\gamma\gamma \to \pi^0\pi^0$ is finite~\cite{Bijnens:1987dc,Donoghue:1988eea}.

The same arguments apply to the QCD axion, which couples to $\pi^+\,\pi^-$ at tree-level in the $\mathcal{O}(p^2)$ Lagrangian, and therefore couples to $\gamma\gamma$ at one loop. In the \textit{s.m.}, we derive the coupling of two axions to two photons, whose size is approximately
\begin{align}
    \nonumber \mathcal{L}_{a^2F^2} &\simeq \frac{\alpha}{16\pi^2}\,\frac{m_u \, m_d}{(m_u + m_d)^2} \frac{\pi}{3} \left(\frac{a}{f_a}\right)^2 \,F_{\mu\nu}F^{\mu\nu} + \mathcal{O}(p^6) \\
    & \simeq \frac{\alpha}{16\pi^2}\frac{\pi}{3} \frac{m_a^2}{\epsilon \, m_\pi^2\, f_\pi^2}\,a^2\,F_{\mu\nu}F^{\mu\nu} + \mathcal{O}(p^6)\ .
    \label{eq:QCDquad}
\end{align}

We identify $\cFF = \pi\,m_u\,m_d/3\,(m_u+m_d)^2\sim 0.2$ when comparing with the form of Eq.~\eqref{eq:QuadAxionPhoton}.
In the second line of Eq.~\eqref{eq:QCDquad} we have written the coupling in terms of the axion mass
\begin{align}
\label{eq:TunedQCDaxMass}
m_a^2 \simeq \epsilon \frac{m_u m_d}{(m_u+m_d)^2}\frac{m_\pi^2 f_\pi^2}{f_a^2} \ ,
\end{align}
where $\epsilon$ encodes possible deviations from the usual QCD prediction~\cite{Hook:2018jle,DiLuzio:2021pxd,Banerjee:2022wzk} and is typically taken to be $\eps \lesssim 1$. We see the expected result that any non-derivative coupling is suppressed by the shift-symmetry breaking parameter, the axion mass, and goes to zero when the shift symmetry is restored.  
Crucially, the denominator has no powers of $f_a$ when the numerator is expressed in terms of $m_a$, and therefore the suppression is not as small as might have been anticipated on dimensional grounds. Indeed, since the operator is generated through the same dynamics as the axion potential at $\Lambda$, the naive power counting should have been that $\cFF \sim (m_\pi f_\pi)^2/\Lambda^4$, which is confirmed in the detailed computation.

Higher-order one-loop and tree-level corrections to Eq.~\eqref{eq:QCDquad} appear at $\mathcal{O}(p^6)$ in the $\xPT$ power-counting scheme, and can safely be neglected.

\begin{figure*}[t]
    \includegraphics[width=1.55\columnwidth]{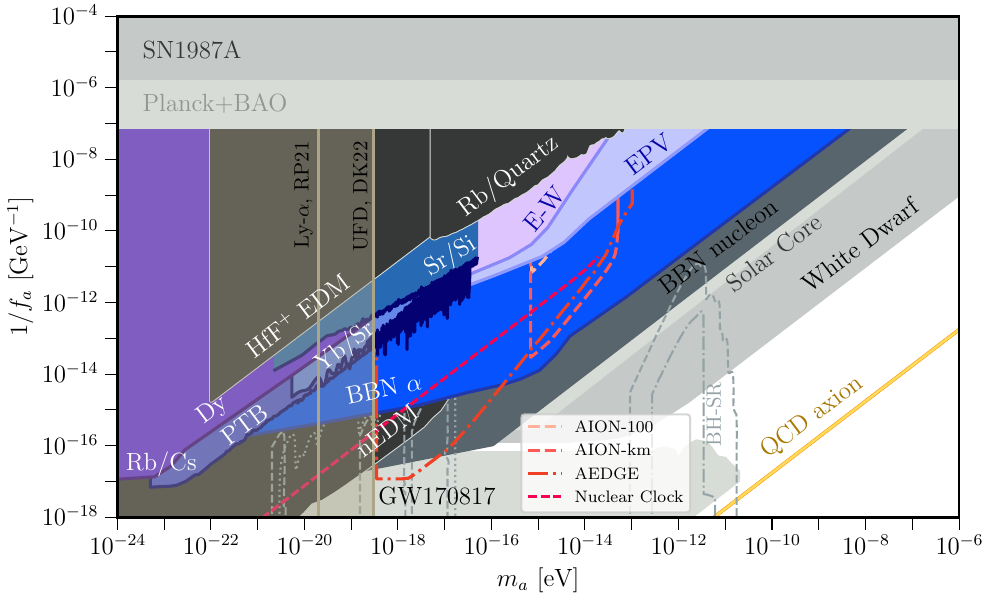}
    \caption{New constraints on the decay constant $f_a$ as a function of the mass $m_a$ for axion dark matter, shown in various shades of blue and purple. New constraints from 
    atomic clocks are shown~\cite{VanTilburg:2015oza,Zhang:2022ewz,Hees:2016gop,PhysRevLett.125.201302,Filzinger:2023zrs,Sherrill:2023zah}, as well as from E\"ot-Wash~\cite{Wagner:2012ui,Hees:2018fpg} and MICROSCOPE~\cite{Berge:2017ovy,Hees:2018fpg} searching for fifth forces and violations of the equivalence principle respectively. Finally, new constraints from BBN~\cite{Bouley:2022eer} 
    are shown. In addition to new constraints, we show projections for future atom interferometer experiments AION-100/MAGIS, AION-km and AEDGE~\cite{Badurina:2019hst,Abe_2021,Badurina:2021rgt}, as well as from a Nuclear clock~\cite{Banerjee:2020kww} with sensitivity $|\delta \alpha|/\alpha = 10^{-22}$. Also shown are existing constraints on tuned QCD axions, such as searches for EDMs ($\text{HfF}^+$~\cite{PhysRevLett.126.171301} and $\text{n}$~\cite{Abel:2017rtm}), Rb clocks~\cite{Zhang:2022ewz}, BBN from to the coupling to nucleons~\cite{Blum:2014vsa}, in-medium effects on the tuned QCD axion potential from the Sun~\cite{Hook:2017psm} and White Dwarfs~\cite{Balkin:2022qer}, SN1987A~\cite{Lucente:2022vuo}, cosmology~\cite{Caloni:2022uya}, and from GW170817~\cite{Zhang:2021mks}. We also show exclusions from black hole superradiance~\cite{Mehta:2020kwu,Unal:2020jiy,Baryakhtar:2020gao} as dashed grey lines. Analysis of ultra-faint dwarf (UFD) galaxies~\cite{PhysRevD.106.063517} and of the Lyman-$\alpha$ forest~\cite{PhysRevLett.126.071302} exclude wave-like DM with very low masses. The existing constraints in dark grey shades indicate that they assume the axion is dark matter, while the lighter shades of grey do not rely on axions as dark matter.}
    \label{fig:QCDplot}
\end{figure*}

\vspace{3pt}
\emph{Axion-like Particles.} --- 
\label{sec:ALPs}
ALPs are often characterised as possessing a mass $m_a$ and decay constant $f_a$ that are unrelated. This is a convenient way of considering the phenomenology of ALPs as an effective field theory (EFT) while setting aside unknown UV dynamics. However, given this ignorance of the UV, one must be careful about consistently building the EFT and including all possible operators (for recent discussions see~\cite{Song:2023lxf,Grojean:2023tsd}). As we saw in the preceding discussion of the QCD axion, the dynamics that breaks the axion shift symmetry also generates the quadratic axion-photon coupling. Similar arguments can be applied to an ALP. 

A simple QCD-like model for an ALP with a quadratic coupling to photons is an $SU(N)\otimes U(1)'$ sector where $SU(N)$ instantons break the ALP shift symmetry, and there are chiral fermions charged under both $SU(N)$ and $U(1)'$. If the chiral fermion masses are $\mathcal{O}(\text{GeV})$, they can have an effective charge under EM of $\qeff \lesssim 0.1 e$ through kinetic mixing of the $U(1)'$ with $U(1)_{\rm EM}$. The dynamics of the $SU(N)$ sector ensure that the ALP couples to the kinetic term of the $U(1)'$, while the kinetic mixing induces a corresponding coupling to $U(1)_{\rm EM}$ with a suppression from the effective charge. The resulting quadratic ALP-photon operator assuming $N_f=2$ with degenerate $SU(N)$ quark masses is
\begin{align}
    \label{eq:ALPquad}
    \mathcal{L}_{a^2F^2} &\simeq \frac{(\qeff)^2\,\alpha}{16\pi^2} \frac{\pi}{12}\left(\frac{a}{f_a}\right)^2\,F_{\mu\nu}F^{\mu\nu} \ .
\end{align}
We can relate this coupling to the ALP mass as in the case of the QCD axion, with $m_a^2 f_a^2 \simeq \epsilon_{\rm ALP} \, m_{\pi'}^2\, f_{\pi'}^2$. The scale $\Lambda' \sim 4\pi f_{\pi'}$ of the $SU(N)$ sector must be sufficiently heavy compared with the light quark mass scale, such that the price to pay for having a light ALP is that $\epsilon_{\rm ALP}$ must be very small. Explicit computation in the \textit{s.m.} shows that for this construction of the ALP-photon coupling, we have $\cFF \simeq \qeff^2(\pi/12)$ which can be $\mathcal{O}(10^{-2})$ for $\Lambda' \sim \text{TeV}$.

An alternative construction of the quadratic operator starts from a UV Lagrangian in which the complex scalar field containing the radial ($
\rho$) and ALP fields couples to fermions charged under $U(1)_{\rm EM}$, similar to the KSVZ model~\cite{Kim:1979if,Shifman:1979if}.\footnote{A DFSZ-like model~\cite{Dine:1981rt,Zhitnitsky:1980tq} would result in tree-level couplings to QCD.} Without an explicit shift-symmetry breaking operator, no quadratic coupling of the ALP to photons is generated upon integrating out the fermions. However, an operator of the form $(\rho/f_a) FF$ \emph{is} generated. Since the radial mode mass is $M_\rho \sim \mathcal{O}(f_a)$, one might think this operator is never relevant for the ALP. However, the potential typically contains a term of the form $V(\rho,a) \supset S[a]\,\rho + \text{h.c.}$ such that upon integrating out $\rho$, the operator $(\rho/f_a) FF \to (S[a]/f_a M_\rho^2) FF$. For the canonical potential with no symmetry breaking, $S[a] \sim (\pd a)^2/f_a$, so that the original intuition that the first quadratic axion-photon operator is $\mathcal{O}((\pd a)^2/f_a^4)$ appears to be confirmed. However, if the UV does not respect the full ALP shift symmetry but only the milder $a\to a+ 2n\pi f_a$ $\mathbb{Z}_n$ symmetry, e.g., $S[a] \sim g^2 f_a \cos(a/f_a)$ with $g$ a dimensionful parameter, integrating out $\rho$ leads to an operator $\sim(g^2 a^2/f_a^2 M_{\rho}^2) F F$. A precise calculation is given in the \textit{s.m.}, yielding $\cFF = (4\pi/3)Q^2 (g/M_\rho)^2$, where $Q$ is the charge of the fermions integrated out in the UV. While the potential we give in the \textit{s.m.} does not lead to a new contribution to the ALP mass, the symmetry-breaking removes some of the protection of the small mass. 
Significant tuning could therefore be required for the ALP mass to remain small in the IR.

In a sense, the two constructions above reflect the same overarching result: dynamics that breaks the full axion shift symmetry (possibly to the smaller $\mathbb{Z}_n$ symmetry) at a certain scale leads to an $a^2F^2$ operator with a coefficient given by a ratio of some power of the shift-breaking parameter over the scale of the breaking. For the QCD-like model, this ratio is $\sim(m_\pi f_{\pi})^2/\Lambda^4\sim 1$, while for the UV-driven model this ratio is $(g/M_\rho)^2$.

\begin{figure*}
    \includegraphics[width = 1.55\columnwidth]{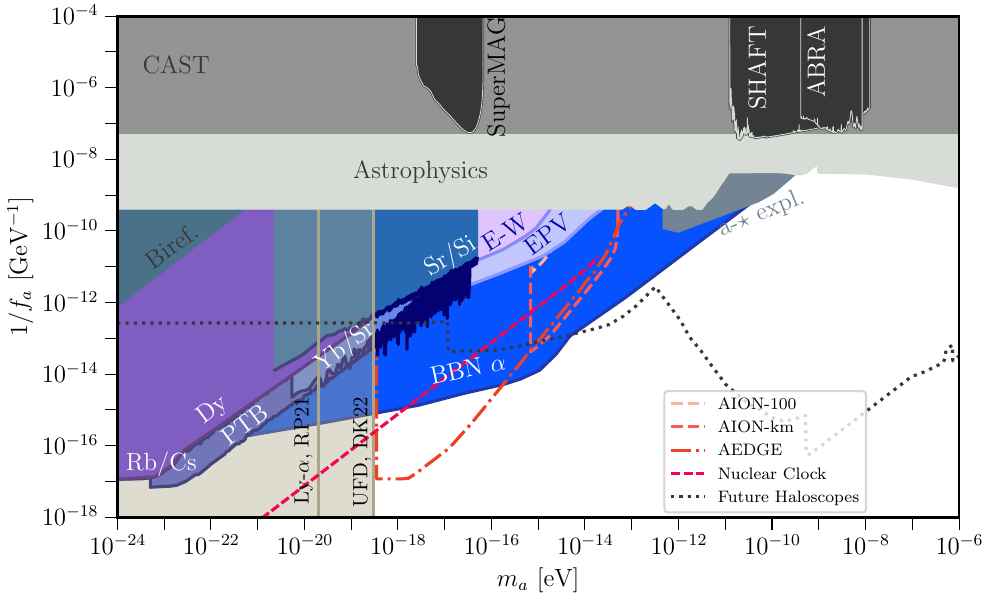}
    \caption{ALP parameter space excluded by quadratic ALP-photon coupling. The colour scheme is the same as in Fig.~\ref{fig:QCDplot}. We have compared with existing and future constraints on linear ALP-photon couplings as discussed in the main text. 
    We have set $\cFF = 0.2$, and constraints from variations of $\alpha$ scale as $(\cFF)^{1/2}f_a$.
    Existing constraints on the linear coupling to photons shown here are from CAST~\cite{CAST:2017uph}, Astrophysics~\cite{Reynes:2021bpe,Noordhuis:2022ljw}, Birefringence~\cite{Fedderke:2019ajk,Liu:2021zlt,BICEPKeck:2021sbt, Castillo:2022zfl,SPT-3G:2022ods, POLARBEAR:2023ric}, SHAFT~\cite{Gramolin:2020ict}, ABRA~\cite{Ouellet:2018beu,Salemi:2021gck} and axion star explosions~\cite{Escudero:2023vgv}. Future haloscopes aimed at ALP dark matter in this mass range include~\cite{Stern:2016bbw, Alesini:2019nzq, DMRadio:2022pkf, Berlin:2019ahk,Berlin:2020vrk,Obata:2018vvr,Bourhill:2022alm}. Analysis of ultra-faint dwarf (UFD) galaxies~\cite{PhysRevD.106.063517} and of the Lyman-$\alpha$ forest~\cite{PhysRevLett.126.071302} exclude wave-like DM with very low masses. To compare with constraints on the linear coupling, we use $\gyy \equiv \alpha/(2\pi f_a)$.}
    \label{fig:ALPplot}
\end{figure*}


\vspace{3pt}
\textbf{\emph{Phenomenology of the Quadratic Coupling.}} --- 
The quadratic axion-photon coupling leads to a shift in the fine-structure constant in the presence of a non-zero background field value of the axion. For dark matter axions near a spherically-symmetric, homogeneous body of mass $M$ and radius $R$ with dilaton charge $Q_e$~\cite{Damour:2010rp}, the background field value is~\cite{Hees:2018fpg}
\begin{align}
    \label{eq:axionField}
    a(t) &\simeq \frac{\sqrt{2 \rhoDM}}{m_a} \cos (m_a t + \varphi)X(r)  \ ,\\ 
    X(r) &= \left( 1- s_C(Q_e) \frac{\cFF\alpha\,M}{16\,\pi^3f_a^2}\frac{1}{r}\right) \ .
\end{align}
The function $s_C(Q_e) \sim Q_e\,\text{min}\left[1,\,3/x^2\right]$, accounts for the screening of the scalar near the macroscopic object, and  
\begin{align}
    \label{eq:screeningVariable}
    x = \sqrt{3 Q_e \frac{\cFF\alpha\,M }{16\,\pi^3f_a^2\,R}} \ .
\end{align}
The resulting shift in $\alpha$ is given by
\begin{align}
\label{eq:AlphaShiftDM}
    \frac{\Delta \alpha}{\alpha} \simeq \cFF \frac{\alpha}{4\pi^2}\frac{2 \rhoDM}{m_a^2\,f_a^2} \cos^2 (m_a t + \varphi) X(r)^2 \ ,
\end{align}
with $\varphi$ an arbitrary phase.
The form of Eq.~\eqref{eq:AlphaShiftDM} implies that there is a static shift in $\alpha$, since $\langle \cos^2 x\rangle = 1/2$, and that the time-varying part oscillates at a frequency $\w \simeq 2 m_a$. 

The constraints from a quadratic scalar-photon coupling have previously been considered in, e.g., Refs.~\cite{Olive:2007aj,Stadnik:2015kia, Banerjee:2022sqg, Bouley:2022eer,Brax:2023udt}. There is a far more extensive literature considering a linear scalar-photon coupling, which has recently been summarised in Ref.~\cite{Antypas:2022asj}. To facilitate comparison between constraints on $f_a$ arising from the $a^2F^2$ coupling and constraints on the linear axion-photon coupling, Eq.~\eqref{eq:AxionPhoton}, we use $\gyy \equiv \alpha/2\pi f_a$.

\vspace{3pt}
\emph{Big Bang Nucleosynthesis (BBN).} --- Variations of the fine-structure constant would impact the predictions of standard BBN, as has been discussed previously~\cite{Stadnik:2015kia,Coc:2006sx,Olive:2007aj,Bouley:2022eer}. 
The most sensitive BBN observable is the yield of $^4\text{He}$, measured to be $Y_p^{\rm exp} (^4 \text{He}) = 0.245 \pm0.003$~\cite{ParticleDataGroup:2022pth}, which agrees extremely well with the theoretical prediction in the Standard Model, $Y_p^{\rm th} (^4 \text{He}) = 0.2467 \pm0.0002$~\cite{Yeh:2023nve}.
A careful analysis of the impact of a quadratically-coupled ultralight scalar DM candidate on BBN was recently performed in Ref.~\cite{Bouley:2022eer}, which we recast as a limit on axions through Eq.~\eqref{eq:AlphaShiftDM} with $X(r)=1$. We take the ``zero-T'' result from their analysis, but caution that the true constraint on $f_a$ could be be up to a factor of $\sim 3$ weaker due to thermal loops contributing to the mass of the axion~\cite{Bouley:2022eer}.

For the QCD axion, the resulting constraints are weaker than those arising from the axion-nucleon coupling~\cite{Blum:2014vsa}. This is expected, since the nucleon coupling appears at tree level, while the photon coupling is a one-loop effect. The EM effect should translate into constraints on $f_a$ that are a factor $\sim 4\pi/\sqrt{\alpha} \sim 10^2$ weaker than the nucleon coupling, an estimate which is confirmed in Fig.~\ref{fig:QCDplot}. For an ALP, however, the nucleon coupling is model-dependent such that the BBN $\alpha$ constraint might be the most stringent in much of the ALP parameter space, as seen in Fig.~\ref{fig:ALPplot}.

\vspace{3pt}
\emph{Fifth Forces and the Weak Equivalence Principle.} --- The effects of ultralight scalar dark matter quadratically-coupled to photons on searches for fifth forces and violations of the weak equivalence principle were considered in Ref.~\cite{Hees:2018fpg}. These results apply to the quadratically-coupled axion also, and therefore appear in Figs.~\ref{fig:QCDplot} and~\ref{fig:ALPplot}. The strongest constraints are from the MICROSCOPE experiment~\cite{Berge:2017ovy} searching for violations of the weak equivalence principle, and the E\"ot-Wash torsion balance experiments~\cite{Wagner:2012ui}.

\vspace{3pt}
\emph{Ultra-light Dark Matter Searches.} --- Experiments looking for ultra-light \emph{scalar} dark matter with a coupling to photons are sensitive to the resulting shift in $\alpha$. The axion-induced oscillation of $\alpha$ leads to a change in atomic energy transitions, allowing strong constraints from very precise clocks~\cite{VanTilburg:2015oza,Zhang:2022ewz,Hees:2016gop,PhysRevLett.125.201302,Filzinger:2023zrs,Sherrill:2023zah}. 
These constraints are shown in Figs.~\ref{fig:QCDplot} and~\ref{fig:ALPplot}. 
We also show projections for atomic interferometers AION and AEDGE~\cite{Badurina:2019hst},\footnote{In Figs.~\ref{fig:QCDplot},~\ref{fig:ALPplot} we have re-interpreted Fig.~4 of Ref.~\cite{Badurina:2019hst}. However, as shown in Ref.~\cite{Badurina:2021lwr}, the sensitivity of atom interferometers to a scalar with linear couplings is likely slightly weaker. A similar conclusion is expected to hold for the quadratic coupling.} and from a nuclear clock~\cite{Banerjee:2020kww}. Notice that some constraints have abrupt endpoints, ranging from $m_a \sim 10^{-17}\,\text{eV}$ for Dy clocks to $m_a \sim 10^{-13}\,\text{eV}$ for AION-km. This is a result of screening by the Earth~\cite{Hees:2018fpg,Banerjee:2022sqg}, which occurs for $f_a \lesssim (\cFF)^{1/2} \times 10^{11}\,\text{GeV}$, as can be computed from Eqs.~\eqref{eq:axionField},~\eqref{eq:screeningVariable}.

\vspace{3pt}
\emph{Other Phenomenology.} --- 
A quadratic axion-photon coupling can have profound implications for experiments looking for axion DM on Earth due to the screening of the axion field at large coupling to matter. On Earth, the screening effect reduces the amplitude of the axion field drastically if $f_a$ is too small. This affects not only the observables associated to the quadratic axion coupling, but the linear axion couplings as well. The full extent of the implications for existing and planned experiments will be explored in separate work~\cite{ScreeningPaper}.

It has recently been shown that the polar cap regions of neutron stars (NSs) have large $\vect{E}\cdot\vect{B}$ and can therefore source non-DM axions~\cite{Prabhu:2021zve,Noordhuis:2022ljw}. The quadratic coupling to $\vect{B}\cdot\vect{B}$ leads to an effective mass for the axion of order 
\begin{align}
\nonumber \w_{a} &\sim \left(\cFF \frac{\alpha}{4\pi^2} \left(\frac{|\vect{B}|}{f_a} \right)^2\right)^{1/2}\\
&\sim 10^{-9}\,\text{eV} \times\left( \frac{|\vect{B}|}{10^{12}\,\text{G}} \right)\,\left( \frac{\gyy}{5\times 10^{-12} \,\text{GeV}^{-1}} \right) \ .
\end{align}
This ``plasma'' mass for the axion coincides with the lower end of the range of bare axion masses to which the analysis of Ref.~\cite{Noordhuis:2022ljw} is sensitive. A careful re-analysis taking into account this effect is therefore motivated. More generally, the plasma mass from the magnetic field around the NS exceeds the bare mass for $\gyy \gtrsim 7\times 10^{-9}\,\GeV^{-1} \times (m_a/\text{$\mu$eV}) \, (10^{12}\,\text{G}/|\vect{B}|)$.

\vspace{3pt}
\textbf{\emph{Conclusion.}} --- The dynamics that endows an axion with a mass, breaking its shift symmetry, also leads to a non-shift-symmetric quadratic coupling of axions to photons. In the case of the QCD axion, we show that the leading contribution to this operator arises at one-loop order. For a generic ALP, some model-building is required, but the quadratic coupling can still be easily generated. The result is that dark matter axions would induce temporal variation of the fine-structure constant $\alpha$, an effect which is severely constrained. In the case of the QCD axion, other constraints are typically stronger, but the quadratic photon coupling offers a new way of independently ruling out significant regions of parameter space. For ALPs, the quadratic photon coupling could be the strongest constraint in wide regions of parameter space, and offers a new way of probing regions that are inaccessible to traditional haloscope searches. Indeed, the existence of a quadratic coupling of axions to matter can have important implications for such searches due to screening near macroscopic objects. A full discussion of these implications will appear in a forthcoming publication~\cite{ScreeningPaper}.

The co-existence of the linear CP-odd and quadratic CP-even couplings of axions could lead to different phenomenology from that of ultralight scalars, which only have CP-even couplings at all orders in the scalar field. A thorough exploration of the implications of this admixture of couplings should be undertaken.

\textit{Acknowledgements.} --- We thank L. Badurina, P. Brax, R. T. D'Agnolo, J. Kley and F. Riva for helpful discussions. The work of CB and SARE was supported by SNF Ambizione grant PZ00P2\_193322, \textit{New frontiers from sub-eV to super-TeV}. The work of JQ is supported by the CNRS IN2P3 Master project A2I. The work of PNHV is supported by the Deutsche Forschungsgemeinschaft (DFG, German Research Foundation) under grant 491245950 and under Germany's Excellence Strategy -- EXC 2121 ``Quantum Universe" -- 390833306.

\bibliographystyle{utphys}
\bibliography{bibliography}

\clearpage
\newpage
\maketitle
\onecolumngrid
\begin{center}
\textbf{\large Quadratic Coupling of the Axion to Photons} \\ 
\vspace{0.05in}
{ \it \large Supplemental Material}\\ 
\vspace{0.05in}
{}
{Carl Beadle, Sebastian A. R. Ellis, J\'er\'emie Quevillon and Pham Ngoc Hoa Vuong}

\end{center}
\setcounter{equation}{0}
\setcounter{figure}{0}
\setcounter{table}{0}
\setcounter{section}{1}
\renewcommand{\theequation}{S\arabic{equation}}
\renewcommand{\thefigure}{S\arabic{figure}}
\renewcommand{\thetable}{S\arabic{table}}
\interfootnotelinepenalty=10000 

\setstretch{1.1}

In this Supplemental Material, we present details of various calculations leading to the main results discussed in the text. In the first section, we show through simple helicity arguments that the linear axion-photon operator of Eq.~\eqref{eq:AxionPhoton} cannot lead to a quadratic operator of the form in Eq.~\eqref{eq:QuadAxionPhoton} at tree level. In the second section, we use Chiral Perturbation Theory ($\xPT$) to show how the operator of Eq.~\eqref{eq:QuadAxionPhoton} \emph{is} generated at one-loop order. Finally, we discuss how this operator could arise in the case of ALPs, both from QCD-like dynamics and from an effective field theory (EFT) perspective.

\section{No tree-level contribution}
\label{sec:TreeHelicity}
We show explicitly why the contribution of the operator $a F_{\mu \nu} \tilde{F}^{\mu \nu}$ to the two-to-two scattering amplitude of axions and photons at tree level is zero.
There will be both $t$- and $u$-channel contributions to the amplitude.
The axion-photon-photon vertex is associated to the structure: 
\begin{align}\label{eq:FeynRuleaFFtildevertex}
        g_{a\gamma \gamma} \epsilon_{\mu \nu \alpha \beta} \varepsilon^{*, \mu}_1 \varepsilon^{*, \nu}_2 p^{\alpha}_1 p^{\beta}_2,
\end{align}
where the subscripts label the distinct outgoing photon momenta and their corresponding polarisation vectors.
We may construct a $4$-point vertex by gluing two of these vertices together, choosing opposite helicities for the outgoing states such that this contributes to the same amplitude as the $a^2 F^2$ operator.
We see that the $t$-channel diagram contribution is: 
\begin{align}\label{eq:tchanneldiagram}
        \mathcal{M}_{t; \; +, -} = \frac{g^2_{a \gamma \gamma} }{t} \left[ \left( \varepsilon^+_1 \cdot \varepsilon^-_2 \right) \left( p_1 \cdot q \right) - \left( \varepsilon^+_1 \cdot q \right) \left( \varepsilon^-_2 \cdot p_1 \right) \right] \left( p_2 \cdot q \right),
\end{align}
where $p_{1,2}$ are the momenta of the outgoing photons and $q$ is the transferred momentum.
One can now easily verify that a gauge choice exists where this is zero.
A nice way to see an appropriate gauge choice is by using the spinor-helicity formalism and the Fierz identities for spinors.
For example, take the following product in Eq.~\eqref{eq:tchanneldiagram}:
\begin{align}\label{eq:gaugechoice2index}
    \varepsilon^-_2 \cdot p_1 &\to \left( - \frac{1}{\sqrt{2}} \frac{\bra{2} \gamma_{\mu} |  s ] }{\left[ 2 s \right]} \right) \left( \frac{1}{2} \bra{1} \gamma^{\mu} | 1 ] \right) \\
    &= - \frac{1}{\sqrt{2}} \frac{\braket{21} \left[ s 1 \right]}{\left[ 2 s \right]}\, .
\end{align}
We have used Fierz identities to go from the first to the second line.
The label `$s$' corresponds to a light-like reference momentum we are free to choose; this encodes our choice of gauge.
Choosing the reference momentum to be that of the other photon's momentum:`$s = 1$', is a good choice as the spinor-helicity brackets are antisymmetric in their arguments, so the product is zero. 
Having chosen $s=1$, the same now occurs for the product of polarisation vectors:
\begin{align}
    \varepsilon_1^+ \cdot \varepsilon_2^- &\to \frac{1}{2} \left( \frac{\bra{r}\gamma_{\mu} | 1 ]}{\braket{r 1}} \frac{\bra{2} \gamma^{\mu} | 1 ]}{[2 1 ]}\right) \\
    &= \frac{\braket{r 2} [11]}{\braket{r1} [2 1]} = 0.
\end{align}
The $u$-channel diagram calculation follows similarly, making the whole contribution zero.
Note that the result could be anticipated by seeing that parity forbids the process
\cite{Bijnens:1987dc}.

\section{Axion-Photon Couplings in Chiral Perturbation Theory}

We derive the non-derivative axion-photon couplings from the Chiral Lagrangian, showing that they first appear at $\mathcal{O}(p^4)$. This is analogous to the couplings between the neutral pions and photons, which also appear at this order~\cite{Bijnens:1987dc,Donoghue:1988eea}. After performing a rotation of the light quark fields to remove the anomaly-induced coupling of the axion to gluons, the axion enters the Chiral Lagrangian through the light quark mass matrix, 
\begin{align}
    M_a = e^{i (a/2f_a)Q_a} \begin{pmatrix} m_u & 0 \\ 0 & m_d \end{pmatrix}e^{i (a/2f_a)Q_a} \ ,
\end{align}
where $Q_a$ is a matrix whose trace is unity, following the notation of Ref.~\cite{GrillidiCortona:2015jxo}. At $\mathcal{O}(p^2)$, this matrix gives rise to the QCD axion mass through the mixing with the neutral pion,
\begin{align}
    \label{eq:p2xPTMass}
    \mathcal{L}_{p^2} = \frac{f_\pi^2}{4}\text{Tr}\left[D_\mu U (D^\mu U)^\dagger \right] +  2 B_0 \frac{f_\pi^2}{4}\text{Tr} \left[ U M_a^\dagger + M_a U^\dagger \right] \ ,
\end{align}
where $U \equiv e^{i \Pi/f_\pi}$, $\Pi = \begin{pmatrix} \pi^0 & \sqrt{2}\pi^+ \\ \sqrt{2}\pi^- & -\pi^0 \end{pmatrix}$ and we define $B_0 \equiv m_\pi^2/(m_u+m_d)$.
Choosing the charge assignment of Ref.~\cite{Georgi:1986df}, $Q_a = M_q^{-1}/\text{Tr}(M_q^{-1})$, which removes the tree-level mixing between the axion and pion, we find the usual relation for the axion-pion potential,
\begin{align}
    V(a) = -m_\pi^2 f_\pi^2 \left( 1- \frac{4 m_u m_d}{(m_u+m_d)^2} \sin^2\left(\frac{a}{2f_a}\right)\right)^{1/2} 
    \ ,
\end{align}
We make this choice in order to simplify the calculation, but note that the final results should be parameterisation-independent~\cite{Bauer:2021wjo}.\footnote{This choice doesn't avoid a kinetic mixing term, but since it is suppressed by $\sim m_a^2/m_\pi^2$, it will be a sub-dominant effect.}

\begin{figure}[h]
    \centering
    \includegraphics[width = 0.7\textwidth]{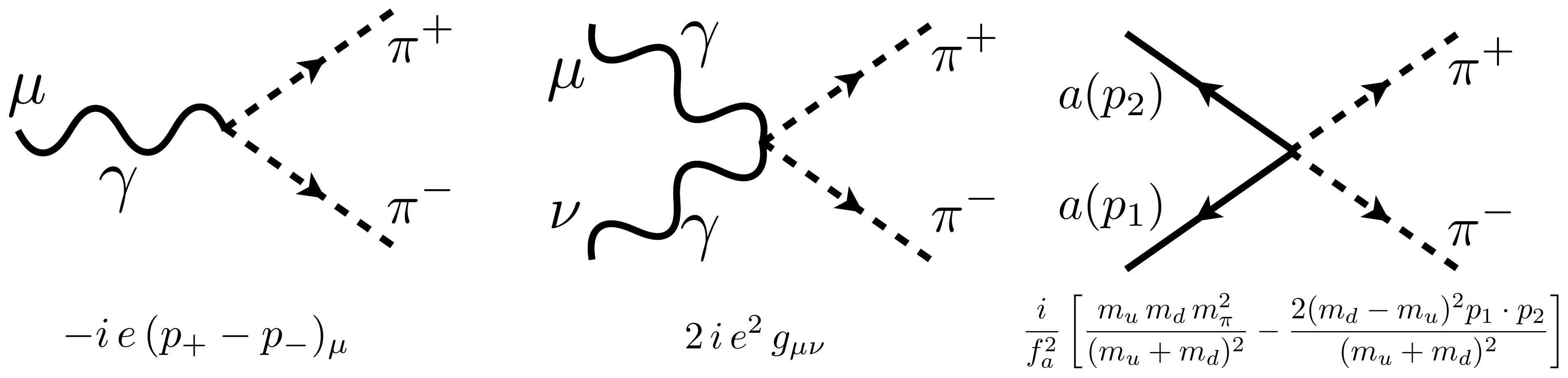}
    \caption{Feynman rules for the vertices from the $\mathcal{O}(p^2)$ chiral Lagrangian leading to a loop-induced $a^2 F^2$ coupling.}
    \label{fig:FeynmanP2}
\end{figure}

\begin{figure}
    \centering
    \includegraphics[width=0.7\textwidth]{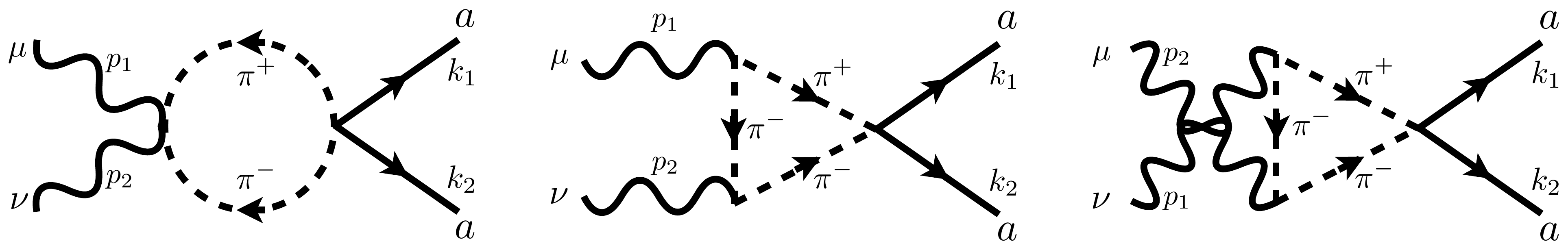}
    \caption{Three diagrams contributing to $\gamma\gamma\to a a$ at $\mathcal{O}(p^4)$ in the $\xPT$ Lagrangian.}
    \label{fig:p4diagrams}
\end{figure}

Expanding Eq.~\eqref{eq:p2xPTMass} to second order in both the axion and the pion fields, we find that it contains terms coupling the charged pions to the photon, which is contained in the covariant derivative $D_\mu$, as well as a term that goes as $a^2\, \pi^+\,\pi^-$. Therefore, at one loop we can construct an $a^2$-photon coupling. The relevant Feynman rules are given in Fig.~\ref{fig:FeynmanP2}, which lead to three one-loop diagrams contributing to the axion-photon coupling shown in Fig.~\ref{fig:p4diagrams}. These loop diagrams, being made of two insertions of $p^2$ operator, are $\mathcal{O}(p^4)$ in the usual $\xPT$ power-counting scheme.

It can be shown that the sum of the three one-loop diagrams of Fig.~\ref{fig:p4diagrams} is finite. Indeed, they must be, since the $\mathcal{O}(p^4)$ $\xPT$ Lagrangian contains no tree-level $a^2F^2$ coupling to absorb any eventual counterterm. 
The amplitude for $\gamma \gamma \to a a$ is
\begin{align}
    &\mathcal{A}(\gamma \gamma \to a a) = \epsilon_\mu(p_1) \epsilon_\nu(p_2) M^{\mu\nu} \ ,
\end{align}
where the tensor multiplying the polarisations is
\begin{align}
    \nonumber i M^{\mu\nu} &= \frac{e^2}{f_a^2\, (m_u+m_d)^2} \left(m_u m_d m_\pi^2 + 2(m_d-m_u)^2 (k_1 \cdot k_2) \right) \\
    &\times \int \frac{d^4 l}{(2\pi)^4} \frac{g^{\mu \nu} (l^2-m_\pi^2) - (2l+p_1)^\mu (2l-p_2)^\nu}{(l^2-m_\pi^2)((l+p_1)^2-m_\pi^2)((l-p_2)^2 - m_\pi^2)} \ .
\end{align}
The possibly divergent terms appear as $g^{\mu\nu} l^2 - 4 l^\mu l^\nu$, and will therefore cancel in four dimensions after dimensional regularisation, leaving a finite result.
Unsurprisingly, this scattering amplitude looks precisely like that of $\gamma \gamma \to \pi^0 \pi^0$, which was calculated long ago~\cite{Bijnens:1987dc,Donoghue:1988eea}, albeit with a different prefactor. The amplitude also has a structure that is manifestly gauge-invariant, and contains a part proportional to the combination $g^{\mu\nu} p_1 \cdot p_2 - p_2^\mu p_1^\nu$, which is characteristic of renormalisation of the coupling $\alpha$:
\begin{align}
    \nonumber i M^{\mu \nu} =& \left(\frac{-i}{8\pi^2} \right)\frac{e^2}{f_a^2\, (m_u+m_d)^2} \left(m_u m_d m_\pi^2 + 2(m_d-m_u)^2 (k_1 \cdot k_2) \right) \, \left( \frac{g^{\mu \nu}(p_1 \cdot p_2)- p_2^\mu p_1^\nu}{p_1\cdot p_2} \right) \\
    &\times \left[1- \frac{2\,m_\pi^2}{s}\left(\text{Li}_2\left(\frac{2\sqrt{s}}{\sqrt{s}-\sqrt{s-4\,m_\pi^2}} \right) + \text{Li}_2\left(\frac{2\sqrt{s}}{\sqrt{s}+\sqrt{s-4\,m_\pi^2}} \right) \right) \right] \, .
\end{align}
In the $s \equiv p_1+p_2 \to 0$ limit, this gives
\begin{align}
    \label{eq:s0p4Coupling}
    M^{\mu\nu} = \left( g^{\mu \nu}(p_1 \cdot p_2)- p_2^\mu p_1^\nu \right) \Pi(0)\ ,\quad \Pi(0) \simeq \frac{e^2\, m_u\, m_d}{48\, \pi^2\,f_a^2 \, (m_u+m_d)^2} \ .
\end{align}
The result is that $\alpha$ is modified from its initial value $\alpha_0$ as
\begin{align}
    \alpha \simeq \alpha_0 \left(1 + \frac{\alpha_0\, m_u\, m_d\, a^2}{12\, \pi\,f_a^2\, (m_u+m_d)^2} \right) \ ,
\end{align}
where $a$ in the case of dark matter axions has a vacuum expectation value, $\langle a(t)^2 \rangle \equiv \rhoDM/m_a^2$, so that the shift will be non-zero.
The same shift in $\alpha$ can be obtained by writing a quadratic axion-photon operator as in Eq.~\eqref{eq:QuadAxionPhoton}.

\section{ALP couplings to Photons}

We discuss the generation of the quadratic coupling of an ALP to photons, fleshing out the arguments made in the main text. Two approaches are possible: IR dynamics similar to that which generates the $a^2F^2$ coupling for the QCD axion also apply to the ALP; the ALP shift symmetry is broken, leading to $a^2F^2$ operators being generated by UV dynamics. 
For the quadratic coupling to be phenomenologically relevant, some level of tuning of the ALP mass will likely be required in both cases.

\subsection{QCD-like Dynamics}

A simple model for realising QCD-like dynamics for the ALP-photon coupling consists of a sector with gauge symmetry $SU(N_c) \times U(1)'$. Instantons of the $SU(N_c)$ sector will generate a potential for the ALP,
\begin{align}
    V(a) \sim m^{N_f}\Lambda'^{4-N_f}\cos\left(\frac{a}{f_a}\right) \ ,
\end{align}
where $m$ is the mass scale of the $N_f$ `quarks' which remain light at the condensation scale of the $SU(N_c)$ sector, $\Lambda'$, in analogy with QCD. 
Assuming $N_f=2$, with both fermions having equal mass, we can use $\xPT$ to refine our estimate to
\begin{align}
    V(a) = - m_{\pi'}^2 f_{\pi'}^2 \cos\left(\frac{a}{2f_a}\right) \simeq - \frac{1}{2} m_a^2 a^2 \ , 
\end{align}
with $m_a^2 f_a^2 = m_{\pi'}^2 f_{\pi'}^2/4$. 

Let us now consider the $U(1)'$, which we will take to be unbroken. The couplings of the ALP to the dark photon will have exactly the same structure as those of the QCD axion to photon, so that we have
\begin{align}
    \alpha' \simeq \alpha'_0 \left(1 + \frac{\alpha'_0\, a^2}{48\, \pi\,f_a^2} \right) \ ,
\end{align}
assuming equal light quark masses. In order to transmit this shift in the dark photon gauge coupling to the regular photon, we can invoke a kinetic mixing (KM) $\chi$ of the two photons,
\begin{align}
    \mathcal{L}^K_{\rm \gamma,\gamma'} = -\frac{1}{4} \left( F_{K,\mu\nu} F_K^{\mu\nu} + F'_{K,\mu\nu} {F'_K}^{\mu\nu} -2\chi F_{K,\mu\nu} {F'_K}^{\mu\nu} \right) - e\, j^\mu A_{K,\mu} - e'\, j'^\mu A'_{K,\mu} \ ,
\end{align}
where the subscript $K$ indicates that these quantities are associated to the KM basis, and $j$ and $j'$ denote the SM and hidden sector currents respectively. For a massless dark photon, the KM basis quantities can be rotated as $A_{K,\mu} \to A_{\mu},~A'_{K,\mu} \to A'_{\mu} + \chi A_{\mu}$ to leading order in $\chi$, such that the photon now couples to the dark current $j'^\mu$
\begin{align}
    \mathcal{L}_{\rm \gamma,\gamma'} = -\frac{1}{4} \left( F_{\mu\nu} F^{\mu\nu} + F'_{\mu\nu} F'^{\mu\nu} \right) - A_{\mu}\left( e\, j^\mu + \chi\,e'\, j'^\mu\right)  - e'\, j'^\mu A'_{\mu} +\mathcal{O}(\chi^2)\ .
\end{align}
We see that the dark current couples directly to the photon with a strength $\chi\,e'$. As a result, in the $\xPT$ analysis of the couplings to external vector fields, the covariant derivative acting on the $U'$ contains not only the dark photon $A'$, but also the regular photon $A$. The dark sector therefore has the same Feynman rules as in Fig.~\ref{fig:FeynmanP2}, only with $m_u=m_d$ and $e\to \chi e'$. Therefore, the shift in $e$ is the same as the shift in $e'$, moderated by the KM factor $\chi$,
\begin{align}
    \label{eq:ALPkmShift}
    \alpha \simeq \alpha_0 \left(1 + \chi^2\alpha'_0 \frac{ a^2}{48\, \pi\,f_a^2} \right) \ .
\end{align}

Above we considered a massless dark photon, for which $\chi\sim 1$ is allowed until we account for the dark fermions. The dark pions will have an effective millicharge of $\qeff = \chi e'/e$ under EM, and are therefore subject to constraints from collider searches~\cite{Prinz:1998ua,Davidson:2000hf,ArgoNeuT:2019ckq,Ball:2020dnx,ArguellesDelgado:2021lek} and stellar cooling~\cite{Davidson:2000hf,Chang:2018rso}. The latter require $\chi\lesssim 10^{-15}$ for $e'\sim 1$, which would make the shift in $\alpha$ unobservably small for reasonable values of $f_a$. If the dark pions have masses $m_{\pi'} \gtrsim \text{MeV}\,(\text{GeV})$, then $\chi \lesssim 10^{-4}\,(0.1)$ is allowed, such that the shift in $\alpha$ can be substantial for reasonable $f_a$.

If the dark photon is massive, the rotation to obtain the mass eigenstate basis is different from above, and results in 
\begin{align}
    \mathcal{L}_{\rm \gamma,\gamma'} = -\frac{1}{4} \left( F_{\mu\nu} F^{\mu\nu} + F'_{\mu\nu} F'^{\mu\nu} \right) -\frac{1}{2}m_{A'}^2 A'_\mu A'^{\mu}- e\, j^\mu A_{\mu} -  A'_{\mu}\left(e'\, j'^\mu + \chi\, e\,j^\mu \right) +\mathcal{O}(\chi^2)\ .
\end{align}
$A_\mu$ is now an admixture $A_{K,\mu} - \chi A'_{K,\mu}$, so that an ALP-induced shift of $e'$ translates into an ALP-induced shift of $e$ at $\mathcal{O}(\chi)$. This follows from the fact that $e'$ is defined in the KM basis through the dark current interaction, so that the shift in $e'$ due to the ALP can be absorbed by a shift in $A'_{K,\mu}$, which then enters $A_\mu$ at $\mathcal{O}(\chi)$. The resulting shift in $\alpha$ is the same as in Eq.~\eqref{eq:ALPkmShift} above.

\subsection{Shift Symmetry-breaking EFT}

For our EFT analysis, let us consider an ALP with a coupling to vector-like (VL) fermions similar to a KSVZ model~\cite{Kim:1979if,Shifman:1979if}. In order to couple to photons, the fermions should have electric charge. The UV Lagrangian is
\begin{align}
    \mathcal{L}_{\rm UV} &= -\dfrac{1}{4}F_{\mu\nu}F^{\mu\nu} + \bar{\psi}_L i\slashed{D}\psi_L + \bar{\psi}_R i\slashed{D}\psi_R 
    + \big( y\phi\bar{\psi}_L\psi_R + \text{h.c.} \big)
    + \partial_{\mu}\phi^{\dagger}\partial^{\mu}\phi - V(\phi^{\dagger}\phi)
    \, ,
    \label{eq:ShiftInvEFT}
\end{align}
with $\phi$ being the complex scalar field  containing both the radial field and the axion. As written, the Lagrangian is invariant under a global $U(1)$ transformation, and the potential can be written as
\begin{align}
    V(\phi^\dagger \phi) = \lambda\left(\phi^\dagger \phi - \frac{f_a^2}{2} \right)^2 \ .
\end{align}
The field $\phi$ admits two commonly used and equivalent representations, one linear with $\phi_l = \frac{1}{\sqrt{2}}\left(\sigma + f_a + i \alpha \right)$, and one polar with $\phi_p = \frac{1}{\sqrt{2}}\left(\rho + f_a \right)\exp(ia/f_a)$. The polar representation makes the shift symmetry that acts on the axion field $a$ evident, while the linear representation obscures it. There exists a map between the two representations and to leading order, $\rho \sim \sigma$ and $a \sim \alpha$.

The $a^2F^2$ operator is not generated from the Lagrangian of Eq.~\eqref{eq:ShiftInvEFT} upon integrating out the VL fermions $\psi_{L,R}$, as expected given the shift-invariance of the Lagrangian. In order to see how such an operator is generated if the shift-symmetry is broken, it is instructive to examine this result. 

It is straightforward to calculate in the polar representation, where the Yukawa interaction of $\phi$ with the VL fermions is $\mathcal{L} \supset (1/\sqrt{2})\,y (\rho+f_a) e^{ia/fa} \bar{\psi}_L\psi_R + \text{h.c.}$  We can then demonstrate that the coefficient of the operator $a^2F^2$ is zero in two ways. In the first, we can expand the Yukawa interaction in powers of $a/f_a$, leading to two diagrams contributing to the operator: a box containing two vertices linear in $a$ and a triangle containing one vertex quadratic in $a$. Integrating out $\psi$ using the universal structures of the fermionic Universal One-Loop Effective Action (UOLEA) of Ref.~\cite{Ellis:2020ivx} we obtain
\begin{align}
    \mathcal{L}_{a^2F^2}^{1\text{-loop}} &= \dfrac{i^2}{16\pi^2}\dfrac{1}{3M_\psi^2}\bigg[M_\psi^2\dfrac{a^2}{f_a^2}\bigg](iQ_\psi e)^2F_{\mu\nu}F^{\mu\nu}
    + \dfrac{i^2}{16\pi^2}\dfrac{2}{3M_\psi}\bigg[M_\psi\dfrac{(ia)^2}{2f_a^2} \bigg](iQ_\psi e)^2F_{\mu\nu}F^{\mu\nu}
    = 0 \ ,
    \label{La2F2: 1-loop quadratic exp}
\end{align}
where $Q_\psi$ is the EM charge of the VL fermion in units of $e$, and $M_\psi = y\,f_a/\sqrt{2}$.
The first term in Eq.~\eqref{La2F2: 1-loop quadratic exp} corresponds to the box diagram, while the second corresponds to the triangle. They precisely cancel each other, as indeed they should. A more elegant way of obtaining the same result is to make use of the symmetries of the UV Lagrangian, and perform a chiral rotation of the fermion field $\psi \to \exp(-i a \gamma_5/2f_a) \psi$ to remove the $a$-dependent phase in the Yukawa interaction in favour of a manifestly shift-symmetric derivative coupling $\mathcal{L} \supset -\frac{\pd_\mu a}{2 f_a}\bar{\psi} \gamma^\mu \gamma_5 \psi$. Using the fermionic UOLEA we find that the coefficient of the operator $\mathcal{O}((\pd a)^2 A^2)$, which can map to $a^2F^2$, is zero as expected. Both approaches demonstrate that the symmetry structure of the Lagrangian is responsible for ensuring that symmetry-breaking operators are not generated.

In the above analysis, we have neglected the radial mode $\rho$. It has a linear coupling to the fermions, such that upon integrating them out, we obtain a $\rho F^2/f_a$ operator with a non-zero Wilson coefficient. Integrating out the $\rho$ at tree level, from the Lagrangian of Eq.~\eqref{eq:ShiftInvEFT} we find that this leads to an operator $\sim(\pd a)^2 F^2/f_a^4$, since the classical background field value of $\rho$ is $\rho_c \simeq (\pd_\mu a)^2/(f_a\,M_\rho^2)$, with $M_\rho^2 = 2\lambda f_a^2$. However, this also means that if the potential for $\phi$ contains non-shift-symmetric terms, $\rho_c$ could be the origin of an $a^2F^2$ operator. For example, adding a shift-symmetry-breaking potential that preserves CP and a $\mathbb{Z}_n$ symmetry for $a$,
\begin{align}
    V(a,\rho)_{\rm s.b.} = g^2 \left(\phi^\dagger \phi-\frac{f_a^2}{2} \right) \left(1-\cos\left(\frac{a}{f_a}\right) \right) \ ,
\end{align}
where $g$ is a dimensionful parameter, leads to $\rho_c = (\pd_\mu a)^2/(f_a\,M_\rho^2) + a^2 g^2/2 f_a M_\rho^2$, and therefore both the $(\pd a)^2 F^2/f_a^4$ and $a^2F^2$ operators. We do not specify the origin of this potential, but merely point out that it is possible to generate the $a^2F^2$ coupling without generating a mass for $a$, without violating CP, and retaining the residual $\mathbb{Z}_n$ symmetry for $a$.

Integrating out first the fermions with the UOLEA, and then integrating $\rho$ out by setting it to its classical background field value, we find
\begin{align}
    \mathcal{L}_{a^2 F^2}^{\rm 1-loop} &\supset \dfrac{i^2}{16\pi^2}\dfrac{2}{3M_{\psi}}\left[ M_{\psi}\dfrac{\rho}{f_a} \right](iQ_{\psi}e)^2F_{\mu\nu}F^{\mu\nu} \bigg|_{\rho = \rho_c(a)}
    \\
    &= \dfrac{1}{48\pi^2}(Q_{\psi}e)^2\dfrac{g^2}{f_a^2M_{\rho}^2}a^2F_{\mu\nu}F^{\mu\nu} + \dfrac{1}{24\pi^2}(Q_{\psi}e)^2\dfrac{1}{f_a^2M_{\rho}^2}(\partial_{\mu} a)^2F_{\nu\sigma}F^{\nu\sigma}
    \, .
\end{align}
In the second line, we have replaced $\rho \equiv \rho_c = \dfrac{(\partial a)^2}{f_a M_{\rho}^2} + \dfrac{a^2 g^2}{2f_a M_{\rho}^2} $. 
Comparing with Eqs.~\eqref{eq:QuadAxionPhoton},~\eqref{eq:AlphaShift} one can identity the value of $c_{F^2}$ and $\alpha(a)$ as,
\begin{align}
    \cFF = \dfrac{4\pi}{3}Q_{\psi}^2\dfrac{g^2}{M_{\rho}^2}
    \, , \quad
    \alpha(a) = \alpha\left(1 + \dfrac{Q_{\psi}^2\alpha}{3\pi}\dfrac{g^2a^2}{M_{\rho}^2f_a^2} \right) \ .
\end{align}

More generally, the condition for the $a^2F^2$ operator to be generated by a symmetry-breaking potential is that the potential take the form
\begin{align}
    V_{\rm s.b.} \supset S[a] \rho + \text{h.c.} \ ,
\end{align}
where we can further impose that $S[a]$ be an even function of $a$ to preserve CP, and that it be a trigonometric function of $a/f_a$ in order for $a$ to possess a residual $\mathbb{Z}_n$ symmetry.  In this case, we should expect the coefficient of the $a^2F^2$ operator and the corresponding shift in alpha to obey
\begin{align}
    \cFF \propto  \dfrac{Q_{\psi}^2\,f_a}{M_{\rho}^2} \,\frac{\pd^2 S[a]}{\pd a^2}
    \, , \quad
    \alpha(a) \sim \alpha\left(1 + Q_{\psi}^2\alpha\dfrac{S[a]}{M_{\rho}^2 f_a} \right) \ ,
\end{align}
where the leading term is $S[a]\propto a^2/f_a^2$.

\end{document}